# Physical-Layer Network Coding for VPN in TDM-PON

Qike Wang, Kam-Hon Tse, Lian-Kuan Chen, *Senior Member, IEEE*, Soung-Chang Liew, *Fellow, IEEE*

*Abstract*—We experimentally demonstrate a novel optical physical-layer network coding (PNC) scheme over time-division multiplexing (TDM) passive optical network (PON). Full-duplex error-free communications between optical network units (ONUs) at 2.5 Gb/s are shown for all-optical virtual private network (VPN) applications. Compared to the conventional half-duplex communications set-up, our scheme can increase the capacity by 100% with power penalty smaller than 3 dB. Synchronization of ONUs is not required for the proposed VPN scheme.

## I. INTRODUCTION

TIME-DIVISION multiplexing (TDM) passive optical network (PON), like E-PON or G-PON, has been widely deployed for delivering access services due to its low cost and broadcast capability [1]. Conventionally, two optical network units (ONUs) in the same PON cannot communicate with each other directly. The inter-ONU traffic must first be sent upstream to the optical line terminate (OLT) and then broadcasted downstream to all ONUs, wasting bandwidth in both directions. To provide private and secure service, as well as to alleviate extra signal processing at the OLT, the all-optical virtual private network (VPN) was recently proposed to enable direct communication among ONUs in the same PON [2-3]. In all-optical VPN, the inter-ONU traffic can be routed at the remote node using a Fiber Bragg Grating (FBG) [2] or dual distribution fibers [3], and then broadcasted optically to all ONUs without occupying upstream and downstream bandwidth resources. However, only unidirectional inter-ONU communication (half-duplex) is allowed in these schemes due to the star coupler's architecture at the remote node. This limits the capacity of all-optical VPN communications.

Network coding was originally proposed to increase network capacity [4] and has recently been investigated for application in optical networks, including PONs [5-7]. However, all these

The authors are with the Department of Information Engineering, The Chinese University of Hong Kong, Shatin, N. T., Hong Kong SAR, China (e-mail: wqk010@ie.cuhk.edu.hk; tkh008@ie.cuhk.edu.hk; lkchen@ie.cuhk.edu.hk; soung@ie.cuhk.edu.hk). This work is supported in part by GRF 410910 and GRF 414911.

schemes implement the encoding and decoding operation logically after optical-to-electrical conversion. The benefits of network coding at OLT have recently been investigated by numerical simulation [6-7]. Two inter-ONU traffic streams of opposite directions are transmitted to the OLT as upstream in different time slots by dynamic bandwidth allocation (DBA) and are then buffered at OLT electrically. The OLT then encodes the two traffic streams via predefined coding operation (e.g., XOR bits) and then broadcasts the encoded traffic stream to all ONUs as downstream. The respective ONU can decode the right traffic stream by buffering a copy of its own traffic stream. However, it requires large electrical buffer at OLT to store the two inter-ONU traffic streams and occupies additional downstream bandwidth. The coding operation at OLT also increases the workload and consumes additional power. The maximum capacity improvement for inter-ONU communication in these schemes is only 50%.

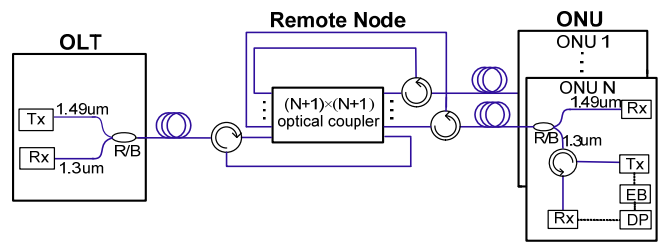

Fig.1 Proposed PNC over TDM-PON architecture for all-optical VPN applications. R/B: red/blue coupler. Tx: transmitter. Rx: receiver. EB: electrical buffer. DP: decoding process. Downstream wavelength: 1.49um. Upstream wavelength: 1.3um.

In this paper, we propose and demonstrate a novel scheme that employs physical-layer network coding (PNC) in all-optical VPN for the first time. The scheme increases the capacity of inter-ONU communication by 100%.

PNC was originally proposed in 2006 as a means to increase throughput in wireless relay networks by implementing the network coding operation directly at the physical layer [8]. When two or more electromagnetic (EM) waves mix together, they add. This addition is a form of network coding realized by nature. Although PNC has been widely studied in wireless networks, its application in optical networks has hardly been explored. In this paper, we experimentally demonstrate that PNC can be applied to an all-optical VPN implementation in TDM-PON. PNC increases the capacity of VPN communication by 100% as well as provides more secure service.

## II. Proposed PNC over TDM-PON Architecture

Fig. 1 depicts the proposed PNC over TDM-PON architecture for all-optical VPN applications. A unique remote node (RN) architecture is proposed, in which the VPN traffic and upstream traffic only experience single-pass splitting loss. At the remote node, each *port 1* of the (*N*+1) three-port optical circulators is connected to one output port of a (*N*+1) × (*N*+1) optical coupler while each *port 3* is connected to one input port of that coupler. One circulator is connected to the OLT while the others are connected to *N* ONUs. The (*N*+1) three-port optical circulators are used such that the optical signal only experiences the high splitting loss caused by optical coupler once. For a 1×32 coupler, it corresponds to ~13-dB reduction in the insertion loss, assuming the insertion loss for the additional circulator is around 1 dB per pass. The conventional all-optical VPN schemes may use a narrow-band FBG in which the optical signal has to pass through the coupler twice [2], or use dual distribution fibers [3], or place a bidirectional amplifier at the OLT [9]. To reduce the cost and power consumption, in the proposed scheme the upstream and VPN communication share the same transmitter at each ONU, whereas [2] has separate transmitters for upstream and VPN communication. To transmit the upstream data and VPN data simultaneously at each ONU, [3] uses subcarrier multiplexer and [9] adopts orthogonal ASK/FSK modulation format, thus increasing the ONU complexity. Like the conventional TDM-PON, in the proposed scheme upstream and downstream can transmit in the same time slot with the use of red/blue coupler. Collisions between the upstream and inter-ONU VPN communications are avoided by DBA protocol at the OLT. If two ONUs need to communicate with each other over the all-optical VPN, the OLT allocates time slots for them and no upstream is transmitted during that time. For the aforementioned VPN schemes, one ONU sends VPN traffic in the first time slot and then the other ONU sends VPN traffic in the second time slot (half-duplex mode). It requires totally two time slots, since all ONUs will receive the VPN traffic due to the star coupler's architecture at the remote node. By employing PNC in the proposed scheme, two ONUs are allowed to transmit optical signals at the same time (full-duplex mode), requiring only one time slot to complete the inter-ONU communication. With that, it increases the capacity of inter-ONU communications by 100% compared to the conventional half-duplex scheme.

The implementation of PNC over TDM-PON for all-optical VPN applications is as follows. For the encoding process (EP), two optical signals are combined together at the remote node. For decoding process (DP), the respective ONU receives the combined optical signal (i.e., its own optical signal and the other ONU's signal), and then converts the optical signal into electrical signal. By subtracting the original copy of the electrical signal buffered previously (i.e., the self-information) from the detected electrical signal, the respective ONU can obtain the other ONU's signal. The EP and DP in this scheme do not require the synchronization between two ONUs, thus greatly enhancing the feasibility of this scheme.

For simplicity, this paper experimentally demonstrates PNC employing the natural property of optical power addition. Therefore, two ONUs should have two different central wavelengths, the short and long wavelengths, to eliminate the interference noise caused by wavelength collision. It is shown in an experiment that a wavelength separation of 0.5 nm is sufficient and there is no requirement on the exact values for the two wavelengths. Low-cost coarse wavelength tuning at ONU can be easily realized by adjusting temperature [10] or injection current [11]. The wavelength-tuning mechanism can be controlled by the OLT using an upper layer protocol. When establishing full-duplex VPN communication between two ONUs, the OLT sends downstream data including temperature or injection current parameters to respective ONUs to select the short or long wavelengths. Although it is possible to filter out the signal from the other ONU by a tunable wavelength filter, this introduces higher cost and requires complex operation of precise wavelength matching. For our proposed PNC system, precise wavelength matching is not needed.

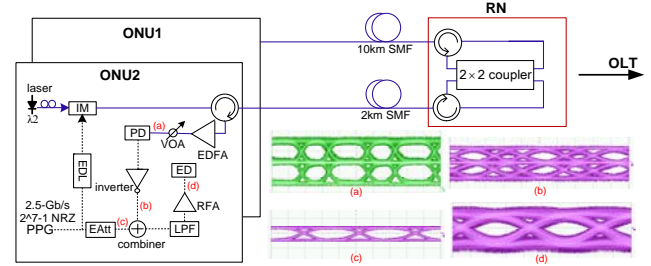

Fig. 2 Experimental setup. Insets show (a) the received two optical signals; (b) the received two electrical signals after photodiode and electrical inverter; (c) the own copy of electrical signal; (d) the decoded electrical signal. IM: intensity modulator; EDL: electrical delay line; EDFA: Erbium doped fiber amplifier; VOA: variable optical attenuator; PD: photodiode; LPF: low pass filter; EAtt: electrical attenuator; RFA: radio frequency amplifier; ED: error detector; RN: remote node.

## III. Experiment and Results

Fig. 2 shows the experimental setup for the full-duplex all-optical VPN inter-ONU communication that employs PNC. Since we only aimed to demonstrate the inter-ONU VPN communication that employs PNC, the OLT was not included in the experimental verification. At ONU2, a CW light at 1548.73nm was intensity-modulated by a 2.5-Gb/s $2^7$-1 pseudorandom binary sequence (PRBS) not-return-zero (NRZ) data before electrical delay line (EDL), which was used to adjust time misalignment between two ONUs for investigating the requirement for synchronization. At ONU1, a CW light at 1552.00nm was intensity-modulated by a 2.5-Gb/s $2^7$-1 PRBS NRZ data. The length of the distribution fiber for ONU2 was 2 km, while the length of the distribution fiber for ONU1 was 10 km. 10-km distribution fiber was adopted to study the effect of long distribution fiber in TDM-PON and to investigate the influence of Rayleigh back scattering (RBS) [12]. Two optical signals were combined at the remote node, which consisted of one 2×2 optical coupler and two 3-port optical circulators. Each *port 1* of the two optical circulators was connected to one side of the 2×2 optical coupler, whereas each *port 3* was connected to the other side of that coupler. *Port 2s* of the two optical circulators were connected to ONU1 and ONU2. Hence, the remote node performed as a star coupler, and the combined optical signal was looped back to both ONU1 and ONU2. The erbium doped fiber amplifier (EDFA) and variable optical attenuator (VOA) in this scheme were just for measuring the receiver sensitivity. After passing through EDFA and VOA, the combined optical signal was converted into electrical signal by a 10-Gb/s p-i-n receiver, which was integrated with an electrical inverter. The converted electrical signal was combined with the original copy of own electrical signal buffered using an electrical combiner for the decoding process. The voltage of the buffered electrical signal was adjusted through an electrical attenuator (EAtt) to achieve the self-signal cancellation in the received electrical signal. The decoded electrical signal passed through a 2.34-GHz low pass filter (LPF) and a radio frequency amplifier (RFA), and was then fed into the error detector (ED) for bit-error-rate (BER) test. The inset (a) in Fig. 2 shows the eye diagram of the combined two optical signals with time misalignment, while inset (b) shows the eye diagram before the electrical combiner, demonstrating the scrambling of the original data when two ONUs are transmitting concurrently. The inset (c) is the eye diagram of original electrical signal, and the inset (d) is the decoded signal after passing through LPF and RFA.

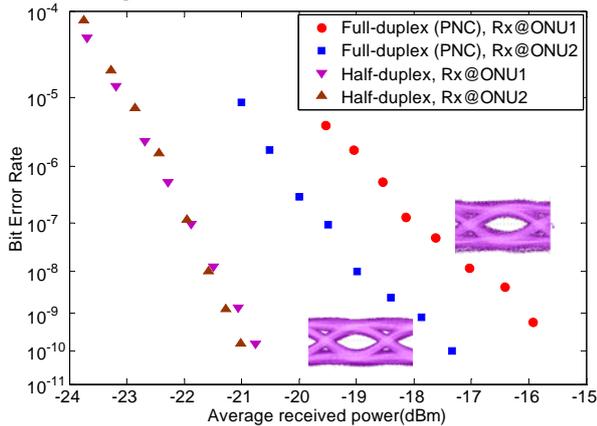

Fig. 3 BER performance comparison of with and without PNC

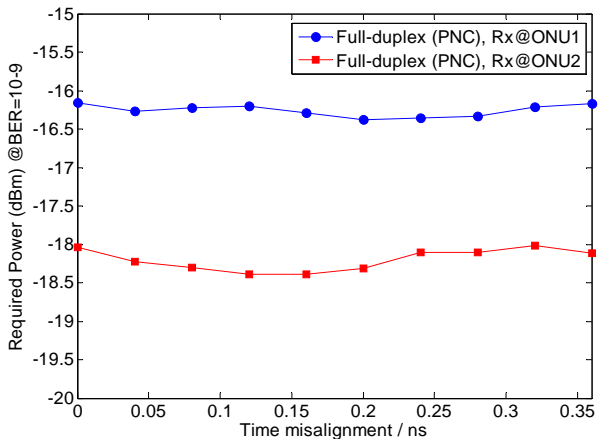

Fig. 4 Required power @BER=$10^{-9}$ versus different time misalignment of two optical signals

We measured the bit error rate (BER) performance of all-optical VPN inter-ONU communications with and without PNC. Fig. 3 shows that ONU2 with 2km-length distribution fiber achieves BER=$10^{-9}$ at about -18 dBm with PNC for VPN communication. The power penalty caused by employing PNC is nearly 3 dB compared to the half-duplex scheme, due to the non-ideal waveform cancellation in the decoding process. The rising and falling edges of electrical signal have changed after electrical-optical-electrical conversion. As the received power decreases, the power penalty becomes smaller because the noise caused by non-ideal waveform cancellation has less influence on the BER performance. However, ONU1 with 10km-length distribution fiber achieves BER=$10^{-9}$ at about -16 dBm with PNC, almost 2 dB worse than that for ONU2 with 2km-length distribution fiber. The 10-km distribution fiber suffers from more severe Rayleigh backscattering, thus degrading the BER performance. Note that most typical distribution fiber has a length less than 5 km in TDM-PON and will experience less degradation from Rayleigh scattering. Fig. 4 shows the required power at BER=$10^{-9}$ versus different timing misalignment of two ONUs' signals. The variation of required power is smaller than 0.4 dB. Thus the synchronization of two ONUs is not required for the proposed scheme, which greatly enhances the feasibility.

## IV. CONCLUSION

For the first time, we experimentally demonstrate an all-optical PNC scheme with error-free full-duplex communication over TDM-PON for VPN applications. By employing PNC, the capacity for inter-ONU communication increases by 100% while the power penalty is no more than 3 dB for 2km-length distribution fiber at BER=$10^{-9}$. A unique RN that uses circulators to reduce the insertion loss of inter-ONU traffic is proposed. We show that the synchronization of two ONUs is not required for the scheme. The proposed scheme can be realized with low-cost simple devices.

REFERENCES

[1] B. Skubic, *et al*., "A comparison of dynamic bandwidth allocation for EPON, GPON, and next-generation TDM PON," *IEEE Communications Magazine,* vol.47, no.3, pp.S40-S48, March 2009
[2] C. J. Chae, *et al*., "A PON system suitable for internetworking optical network units using a fiber Bragg grating on the feeder fiber," *IEEE Photon. Technol. Lett.,* vol. 11, no. 12, pp.1686–1688, Dec. 1999.
[3] N. Nadarajah, *et al*., "Novel schemes for local area network emulation in passive optical networks with RF subcarrier multiplexed customer traffic," *J. Lightw. Technol.,* vol.23, no.10, pp. 2974- 2983, Oct. 2005
[4] S.-Y. R. Li, R. W. Yeung, and N. Cai, "Linear network coding*," IEEE Transactions on Information Theory*, vol. 49, pp. 371–381, 2003.
[5] R.C. Menendez, *et al*., "Efficient, Fault-Tolerant All-Optical Multicast Networks via Network Coding," *OFC/NFOEC* 2008, CA, JThA82
[6] M. Belzner and H. Haunstein, "Network Coding in Passive Optical Networks," In *Proc. ECOC*, Vienna, Austria, Sept. 2009, Paper P6.20
[7] K. Fouli, *et al.,* "Network coding in next-generation passive optical networks*," IEEE Communications Magazine*, vol.49, no.9, pp.38-46, September 2011
[8] S. Zhang, S. C. Liew, and P. P. Lam, "Hot topic: physical layer network coding," in *Proc. 12th MobiCom*, pages 358–365, NY, USA, 2006

[9] Yue Tian, *et al.,* "Demonstration and Scalability Analysis of All-Optical Virtual Private Network in Multiple Passive Optical Networks Using ASK/FSK Format," *IEEE Photon. Technol. Lett.,* vol.19, no.20, pp.1595-1597, 2007

[10] S. Sakano, *et al.,* "Tunable DFB laser with a striped thin-film heater," *IEEE Photon. Technol. Lett.,* vol.4, no.4, pp.321-323, April 1992

[11] Markus Roppelt, *et al.,* "Tuning of an SG-Y Branch Laser for WDM-PON", In *Proc. OFC*, Los Angeles, CA, MAR. 2012

[12] T. H. Wood, et al., " Observation of coherent Rayleigh noise in single-source bidirectional optical fiber systems," *J. Lightw Technol.,* vol. 6, no. 2, pp. 346‑352, Feb. 1988